%% file: geach_euclid_astroph.tex
\documentclass[useAMS,usenatbib]{mn2e}
 \usepackage{times,graphicx,aas_macros,amssymb}

\begin{document}

\title[Predictions of H$\alpha$ number counts]{Empirical H$\alpha$
  emitter count predictions for dark energy
  surveys}\author[J. E. Geach et al.]  {\parbox[h]{\textwidth}{J.\ E.\
    Geach$^1$\thanks{E-mail: j.e.geach@durham.ac.uk}, A. Cimatti$^2$,
    W. Percival$^3$, Y. Wang$^4$, L. Guzzo$^5$, G.~Zamorani$^6$,
    P.~Rosati$^7$, L.~Pozzetti$^6$, A.~Orsi$^1$, C.~M.~Baugh$^1$,
    C.~G.~Lacey$^1$, B.~Garilli$^8$, P.~Franzetti$^8$, J.~R.~Walsh$^7$
    and M.~K\"ummel$^7$}
  \vspace*{6pt}\\
  \noindent $^1$Institute for Computational Cosmology, Department of
  Physics, Durham University, South Road,
  Durham. DH1 3LE. U.K.\\
  \noindent $^2$Dipartimento di Astronomica, Universita di Bologna,
  via Ranzani 1, I--40127, Bologna, Italy\\
  \noindent $^3$Institute of Cosmology and Gravitation, University of
  Portsmouth, Dennis Sciama building, Portsmouth, P01 3FX, UK\\
  \noindent $^4$Homer L. Dodge Department of Physics \& Astronomy, The
  University of Oklahoma, 440
  W. Brooks St., Norman, OK 73019. U.S.A.\\
\noindent $^5$INAF, Osservatorio Astronomico di Brera,
via Bianchi 46, I--23807 Merate (LC), Italy\\ 
\noindent $^6$INAF, Osservatorio Astronomico di Bologna, via Ranzani
1, 40127 Bologna, Italy\\
\noindent $^7$Space Telescope European Co-ordinating Facility,
European Southern Observatory, Karl Schwarzschild Str. 2, D--85748,
Garching bei M\"unchen, Germany \\
\noindent $^8$INAF - IASF Milano, via E. Bassini 15, I--20133, Milan,
Italy}

\date{}

\pagerange{\pageref{firstpage}--\pageref{lastpage}} \pubyear{2008}

\maketitle

\label{firstpage}

\begin{abstract}
  Future galaxy redshift surveys aim to measure cosmological
  quantities from the galaxy power spectrum. A prime example is the
  detection of baryonic acoustic oscillations (BAOs), providing a
  standard ruler to measure the dark energy equation of state, $w(z)$,
  to high precision. The strongest practical limitation for these
  experiments is how quickly accurate redshifts can be measured for
  sufficient galaxies to map the large-scale structure. A promising
  strategy is to target emission-line (i.e.\ star-forming) galaxies at
  high-redshift ($z\sim0.5$--$2$); not only is the space density of
  this population increasing out to $z\sim2$, but also emission-lines
  provide an efficient method of redshift determination. Motivated by
  the prospect of future dark energy surveys targeting H$\alpha$
  emitters at near-infrared wavelengths (i.e.\ $z>0.5$), we use the
  latest empirical data to model the evolution of the H$\alpha$
  luminosity function out to $z\sim2$, and thus provide predictions
  for the abundance of H$\alpha$ emitters for practical limiting
  fluxes. We caution that the estimates presented in this work must be
  tempered by an efficiency factor, $\epsilon$, giving the redshift
  success rate from these potential targets. For a range of practical
  efficiencies and limiting fluxes, we provide an estimate of
  $\bar{n}P_{0.2}$, where $\bar{n}$ is the 3D galaxy number density
  and $P_{0.2}$ is the galaxy power spectrum evaluated at
  $k=0.2\,h\,{\rm Mpc}^{-1}$. Ideal surveys must provide
  $\bar{n}P_{0.2}>1$ in order to balance shot-noise and cosmic
  variance errors. We show that a realistic emission-line survey
  ($\epsilon=0.5$) could achieve $\bar{n}P_{0.2}=1$ out to $z\sim1.5$
  with a limiting flux of $10^{-16}$\,erg\,s$^{-1}$\,cm$^{-2}$. If the
  limiting flux is a factor 5 brighter, then this goal can only be
  achieved out to $z\sim0.5$, highlighting the importance of survey
  depth and efficiency in cosmological redshift surveys.
  \end{abstract}

\begin{table*}
  \caption{Parameters of the luminosity functions used to derive the empirical
    model of H$\alpha$ counts. All Schechter function parameters have
    been corrected to a common fiducial cosmology
    ($H_0=70$\,km\,s$^{-1}$\,Mpc$^{-1}$, $\Omega_{\rm m} = 0.3$, $\Omega
    _{\Lambda} = 0.7$).}
`
\begin{tabular}{@{\extracolsep{\fill}}lcccccl}
  \hline
  & & \multicolumn{3}{c}{Schechter function parameters}\cr
  Reference & $z$ & $\log L^\star {\rm (erg\ s^{-1})}$ & $\log
  \phi^\star {\rm (Mpc^{-3})}$ & $\alpha$ 
  & EW$_0$ (\AA) & Type\cr
  \hline
  Gallego et al.\ (1995) & $<$0.045 & 41.87 & $-2.78$ & $-1.3$  & $>$10 & UCM survey\cr 
  Shioya et al.\ (2008) & 0.24 & $41.94$ & $-2.65$ & $-1.35$  & $>$9 & Narrowband 0.815$\mu$m\cr
  Yan et al.\ (1999) & $1.3\pm0.5$  & $42.83$ & $-2.82$ & $-1.35$  &
  $\sim$10--130 & {\it HST}/NICMOS Grism 1.5$\mu$m\cr
  Geach et al.\ (2008) & $2.23\pm0.03$ & $42.83$ & $-2.84$ & $-1.35$ & $>$12 & Narrowband 2.121$\mu$m\cr
  \hline
\end{tabular}

\end{table*}

\section{Introduction}

One of the greatest challenges the current generation of cosmologists
faces is to understand the physics underlying the apparent
acceleration of the expansion of the Universe (e.g.\ Riess et al.\
1998; Perlmutter et al.\ 1999). Contemporary models favour the
influence of a dark energy that has come to dominate the energy
density of the universe during the last 8\,billion
years. Unfortunately dark energy is outside the realm of the standard
model, and requires new physics to explain. Nevertheless, many
mechanisms have been proposed, and the potential for establishing
which (if any) is correct experimentally, has caused great fervour
amongst the astronomical community over the past decade.  The reward
for investing a large amount of effort into determining the physics of
dark energy is of course a profound advancement of our understanding
of the fundamental nature of the universe.

A range of dark energy models exist (see Peebles \& Ratra\ 2003 and
Copeland et al.\ 2009 for reviews), however the two most prominent
scenarios attribute the accelerating expansion to (a) a `cosmological
constant' ($\Lambda$) analogous to a non-zero quantum mechanical
vacuum energy that has now come to dominate the overall energy density
of the universe (but 120 orders of magnitude smaller than the value
predicted by quantum physics); or (b) a dynamic scalar field
(`quintessence') which varies with both time and space. Both models
require general relativity to hold on cosmological scales. A third
alternative to explain the acceleration is the failure of general
relativity on large scales, such that the gravity theory itself needs
to be modified (e.g.\ Dvali et al.\ 2000). 

One way of distinguishing $\Lambda$ from quintessence is to measure
the evolution of the expansion of the universe, which is controlled by
the dark energy equation of state, $w(z)$; the ratio between dark
energy pressure $P$ and density, $\rho$. For $\Lambda$ models,
$\rho=-P/c^2$ for all time, such that $w(z)=-1$. Detecting a varying
$w(z)$ would be a possible indication for quintessence.  However, if
gravity is modified, such behaviour could be just an indirect effect
of the failure of general relativity.  Such degeneracy between dark
energy and modified gravity can be lifted only by measuring the growth
rate of cosmic structure $f(z)$ (or its integral $G(z)$), which is
governed by the interplay between the strength of gravity and the
expansion rate of the Universe.  Thus, measuring $w(z)$ and $f(z)$ as
a function of redshift represents the most powerful combination of
observational probes to distinguish among competing models (Guzzo et
al.\ 2008, Wang et al.\ 2008). This is the primary goal of current and
future dark energy surveys (Albrecht et al.\ 2009).

Although the accelerated expansion was discovered using observations
of Type Ia supernovae (Riess et al.\ 1998; Perlmutter et al.\ 1999),
results from these observations are now dominated by systematic errors
(e.g. Hicken et al.\ 2009). Future studies of dark energy therefore
aim to exploit different observations and of those proposed, the use
of Baryon Acoustic Oscillations (BAO) as standard rulers appears to
have the lowest level of systematic uncertainty (Albrect et al.\
2006). BAO are a series of peaks and troughs in the power spectrum,
which quantifies the clustering strength of matter as a function of
scale. They occur because primordial cosmological perturbations excite
sound waves in the relativistic plasma of the early universe: when the
plasma breaks down at recombination, the radiation can be observed as
the Cosmic Microwave Background (CMB), while the fluctuations in the
baryonic material give rise to BAO (Silk\ 1968, Peebles \& Yu\ 1970,
Sunyaev \& Zel'dovich\ 1970, Bond \& Efstathiou\ 1984, 1987, Holtzman\
1989). The BAO signal is on large-scales, which are predominantly in
the linear regime today. It is therefore expected that BAO should also
be seen in the galaxy distribution (Goldberg \& Strauss\ 1998,
Meiksin, White \& Peacock\ 1999, Springel et al.\ 2005, Seo \&
Eisenstein\ 2005, White\ 2005, Eisenstein, Seo \& White\ 2007, Kazin
et al.\ 2009), and can be used as a standard ruler, leading to
measurements of the angular diameter distance $D_A(z)$ and the Hubble
expansion rate $H(z)$, and therefore $w(z)$ (Seo \& Eisenstein\ 2003,
Blake \& Glazebrook\ 2003, Hu \& Haiman\ 2003, Wang\ 2006).

The acoustic signature has now been convincingly detected at low
redshift (Percival et al.\ 2001, Cole et al.\ 2005, Eisenstein et al.\
2005, Huetsi\ 2006) using the 2dF Galaxy Redshift Survey (2dFGRS;
Colless et al.\ 2003) and the Sloan Digital Sky Survey (SDSS; York et
al.\ 2000). Further analyses of the SDSS have led to competitive
constraints on cosmological models (Percival et al.\ 2007, Gaztanaga
et al.\ 2009, Percival et al.\ 2009). Ongoing spectroscopic surveys
aiming to use BAO to analyse dark energy include the Baryon
Oscillation Spectroscopic Survey (BOSS; Schlegel, White \& Eisenstein\
2009), the Hobby-Eberly Dark Energy Experiment (HETDEX; Hill et al.\
2008) and the WiggleZ survey (Glazebrook et al.\ 2007). Ongoing
photometric surveys such as the Dark Energy Survey (DES: {\tt
  http://www.darkenergysurvey.org}), the Panoramic Survey Telescope \& Rapid
Response System (Pan-STARRS: {\tt http://pan-starrs.ifa.hawaii.edu}) aim to
find BAO using photometric redshifts.

The power spectrum (or correlation function) of the galaxy
distribution also contains key information on the growth rate of
structure $f(z)$ (Kaiser\ 1987). This produces large-scale motions
towards density maxima, that contribute a peculiar velocity component
to the measured galaxy redshifts used to reconstruct cosmic structure
in 3D.  The net effect is to produce an anisotropy in the power
spectrum that can be measured to extract an estimate of the growth
rate $f(z)$, modulo the bias factor of the galaxies being observed.
The importance of this well-known effect in the context of dark energy
has become evident only in recent times, when redshift surveys of
sufficient size at $z\sim1$ have started to become available (Guzzo et
al.\ 2008).  Thus, a redshift survey of galaxies provides us with the
ability to obtain an estimate of both key probes of cosmic
acceleration, the expansion rate and the growth rate.

In order to reduce shot-noise and cosmic variance in `precision'
measurements of BAO and redshift distortions, the ultimate
observational challenge is to accurately measure a large number (tens
or hundreds of millions) of redshifts for galaxies spread over a
significant interval of cosmic time, spanning the transition from
matter domination to dark energy domination in the universe, and
covering the majority of the extragalactic ($|b|>20^\circ$) sky,\
$\sim$$2\times10^4$ square degrees. Such a survey can be conducted
using a dedicated survey telescopes from a space platform, as proposed
by the Joint Dark Energy Mission (JDEM: {\tt
  http://jdem.gsfc.nasa.gov}) and European Space Agency's Euclid and
SPACE satellite mission concepts ({\tt http://sci.esa.int/euclid};
Cimatti et al.\ 2009). Some of the ongoing and planned future BAO
surveys, such as WiggleZ, will target emission-line galaxies -- i.e.\
generally star-forming galaxies with easily identifiable
redshifts. The goal of this work is to make a prediction for the
abundance of H$\alpha$ emitting galaxies that these dark energy
surveys can expect using the existing empirical evidence of past and
recent H$\alpha$ surveys out to $z\sim2$.

In this work we use empirical data to build a simple phenomenological
model of the evolution of the H$\alpha$ luminosity function (LF) since
$z\sim2$, and therefore predict the number counts of H$\alpha$
emitters in redshift ranges pertinent to future dark energy surveys
(the empirical model can also be used as a fiducial point for
semi-analytic predictions for the abundance of star forming galaxies,
e.g. Baugh et al.\ 2005; Bower et al.\ 2006; Orsi et al.\ 2009 in
prep). In Section 2 we describe the model, list the principal
predictions and draw the reader's attention to some important
caveats. In Section 3 we discuss the implications of the number count
predictions on planned dark energy surveys, and in Section 4 we
comment on the relevance of cosmological surveys in the near-IR from a
terrestrial base. For luminosity estimates, throughout we assume a
fiducial cosmological model of $H_0 = 70$\,km\,s$^{-1}$\,Mpc$^{-1}$,
$\Omega_{\rm m} = 0.3$ and $\Omega_\Lambda = 0.7$.

\begin{figure} \includegraphics[width=0.49\textwidth]{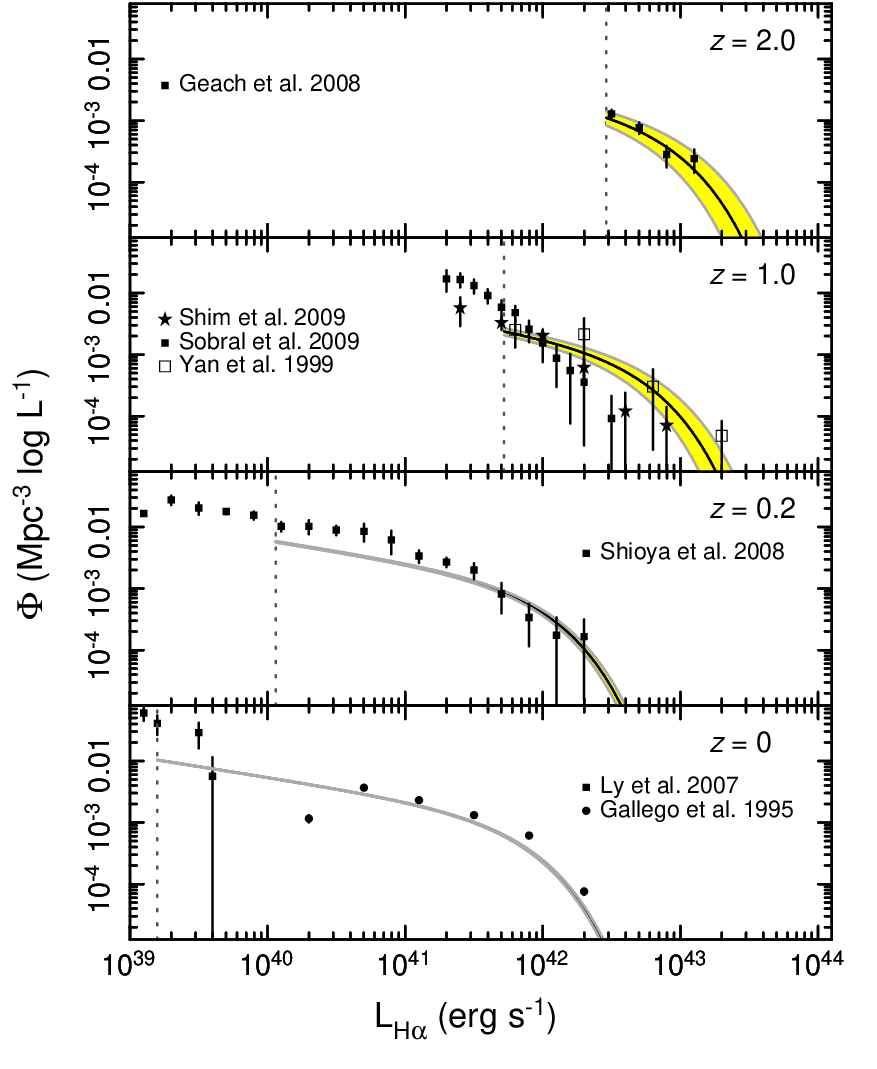}
  \caption{Evolution of the H$\alpha$ luminosity function, assuming
    our simple model of $L^\star\propto(1+z)^Q$ out to $z=1.3$ and no
    evolution to $z<2.2$. The panels show the LF at $z\sim0$, 0.2, 0.9
    \& 2.2, with observational data overlaid (all data has been
    corrected to the same fiducial cosmology used throughout this work
    and {\it not} corrected for extinction). Note that not all of the
    observational data shown here was used to construct the model (see
    \S2.1), however the model is a good representation of the observed
    LFs out to $z\sim2$. The largest discrepancy occurs at $z\sim1$,
    where there is some scatter between different surveys. However, in
    part, this is due to the mixture of survey strategies, and cosmic
    variance in the small fields observed.  The model LFs have been
    truncated at the luminosity limit corresponding to a flux of
    $10^{-16}$\,erg\,s$^{-1}$\,cm$^{-2}$ at each epoch (vertical
    dotted lines). Although there are hints that the faint-end slope
    is steepening out to $z\sim1$, in the flux regime of practical
    interest this does not have a significant impact on our counts
    (also see \S2.2.1 \& Figure\ 3). }
\end{figure}

\begin{figure}
\includegraphics[width=0.49\textwidth]{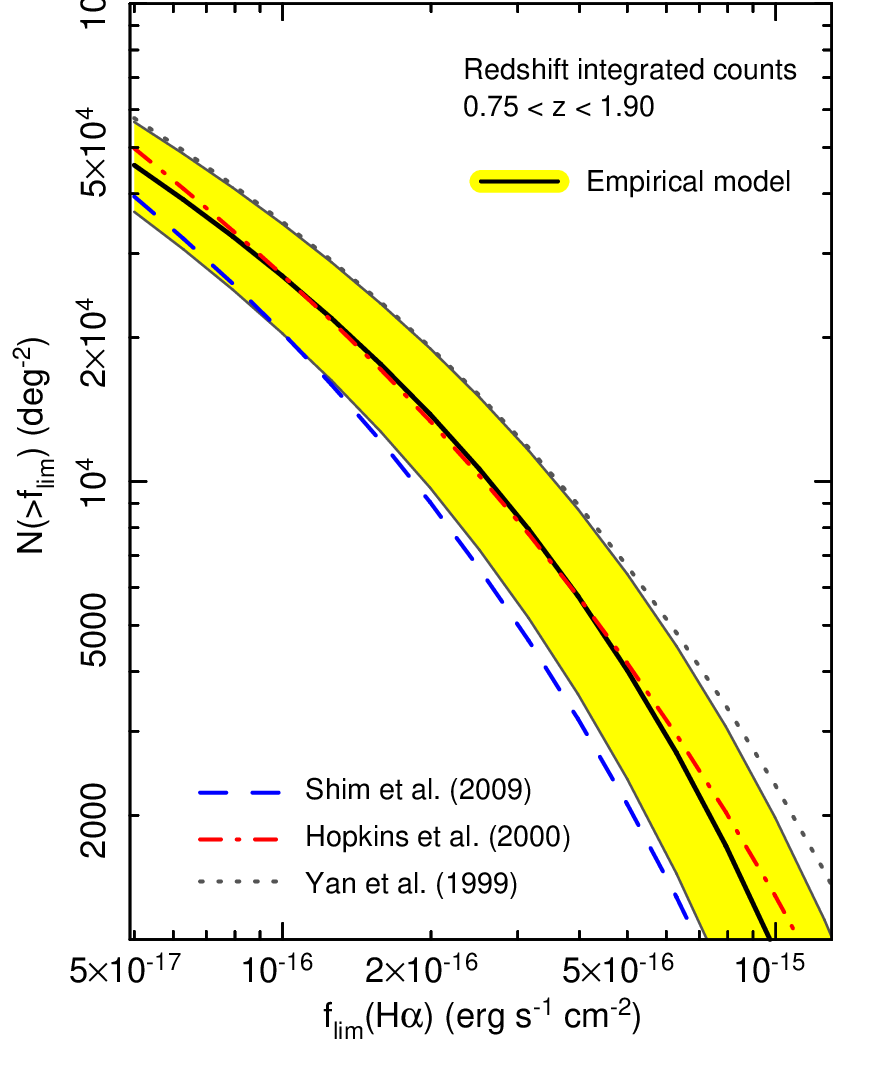}
\caption{A comparison of the predicted number counts of H$\alpha$
  emitters from the simple model to observed counts integrated over
  the redshift range $0.75 < z < 1.90$. The shaded region indicates
  the 1$\sigma$ uncertainty on the model counts. We compare to the
  observational data of the (slitless spectroscopic) surveys of
  McCarthy et al.\ (1999), Hopkins et al.\ (2000) and Shim et al.\
  (2009), where the integrated counts have been calculated from the
  respective luminosity functions, uncorrected for dust extinction (so
  the counts include incompleteness corrections specific to each
  survey). Note that all-sky redshift surveys are unlikely to probe
  below flux limits of $\sim$$10^{-16}$\,erg\,s$^{-1}$\,cm$^{-2}$,
  where uncertainties due to the poorly constrained faint-end slope
  become more important to the count predictions (see \S2.2.1 for more
  details).}
\end{figure}

\begin{figure}
\centerline{\includegraphics[width=0.47\textwidth]{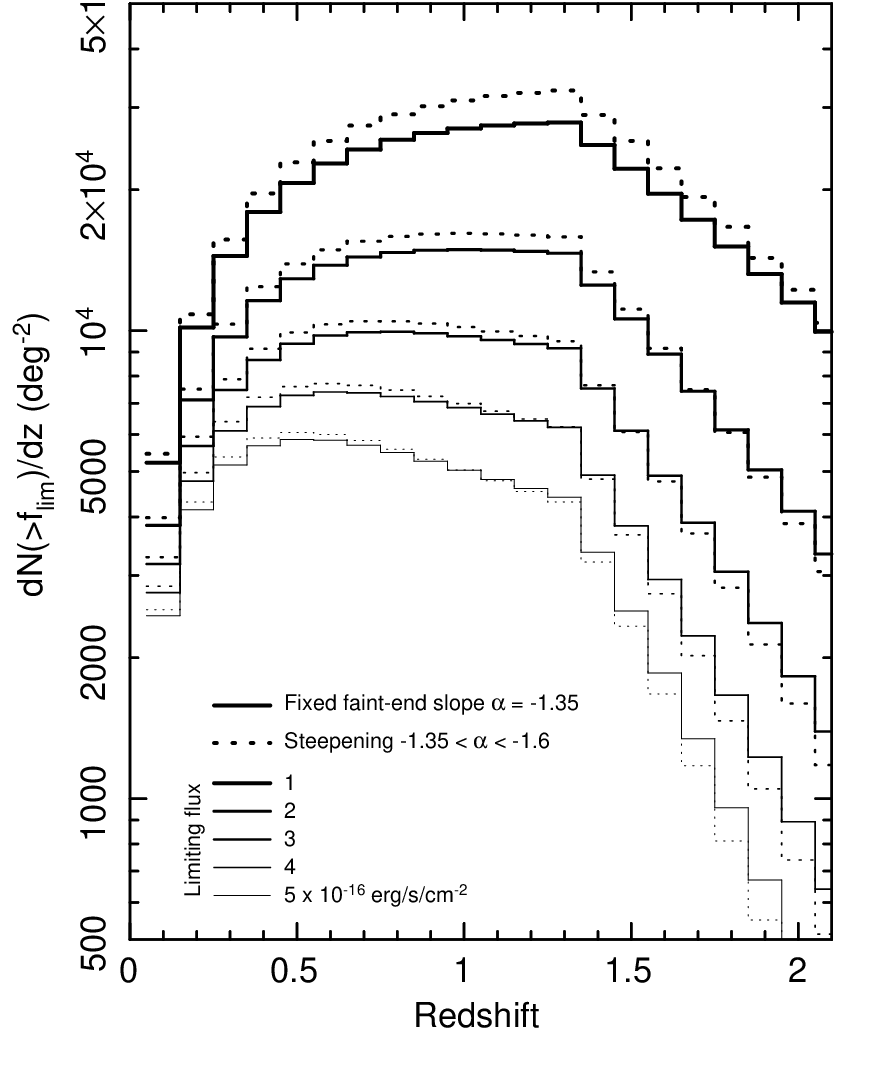}}
\caption{Predicted redshift distribution ${\rm d}N/{\rm d}z$ of
  H$\alpha$ emitters for limiting fluxes of $1$--$5\times
  10^{-16}$\,erg\,s$^{-1}$\,cm$^{-2}$ (thick to thin lines). Note that
  the transition between $L^\star$ evolution and non-evolution at
  $z=1.3$ introduces the sharp fall-off in counts towards
  high-$z$. For comparison, we also show the redshift distribution for
  the same $L^\star$ evolution and fixed $\phi^\star$, but allowing
  the faint end slope to steepen monotonically from $-1.35$ at $z=0$
  to $-1.6$ at $z=2$. The impact this change has on the predicted
  counts in the flux limits of practical interest is negligible, and
  (as expected) more pronounced at fainter limits. }
\end{figure}

\section{A simple model of the evolution of the H$\alpha$ luminosity
  density}

Fortuitously for dark energy surveys, the global volume averaged star
formation rate increases steeply out to $z\sim2$, and flattens (or
perhaps gently declines) towards earlier epochs (e.g.\ Lilly et al.\
1995, Hopkins\ 2004).  This will work in favour of dark energy
surveys, provided the shape of the LF is reasonably well
understood. Locally, star forming galaxies can be easily selected
using the well calibrated and `robust' H$\alpha$ emission line at
$\lambda=6563$\AA\ (e.g. Gallego et al.\ 1995; Ly et al.\ 2007; Shioya
et al.\ 2008). This is a favourable line to target at high-redshift
because it is the least affected by extinction (compared to, say,
[O{\sc ii}]).  The shape of the H$\alpha$ LF in the local Universe is
well characterised, and over the past decade, near-infrared surveys
have tracked the evolution of the LF out to $z\sim2$ (McCarthy et al.\
1999, Yan et al.\ 1999, Hopkins et al.\ 2000; Moorwood et al.\ 2000).
Furthermore, the increasing feasibility of statistically significant
wide-field H$\alpha$ surveys at high-redshift have vastly improved our
picture of how the H$\alpha$ luminosity function has evolved over the
past 8\,Gyr (e.g.\ Geach et al.\ 2008; Shim et al.\ 2009; Sobral et
al.\ 2009).

\subsection{Empirical fit}

Throughout this work, we assume the conventional form of the
luminosity function holds at all epochs -- the Schechter function:
\begin{equation}
\phi(L){\rm d}L = \phi^\star(L/L^\star)^\alpha \exp(-L/L^*){\rm d}(L/L^\star)
\end{equation}
In Table\ 1 we list the Schechter function parameters derived from
four H$\alpha$ surveys spanning $0 < z < 2$, chosen for their
similarity in fitting (all find or fix the faint-end slope
$\alpha=-1.35$) and equivalent width cuts, generally ${\rm
  EW}_0>10$\AA. Note that there is very little evolution in the LF
between $z\sim2$ and $z=1.3$ (Yan et al.\ 1999; Geach et al.\ 2008),
although both of these surveys {\it assume} a fixed faint-end slope of
$-1.35$ similar to that found in the local Universe (necessitated by
the depths of these surveys). In comparison, by $z\sim0$, the
characteristic luminosity $L^\star$ has dropped by an order of
magnitude (Gallego et al.\ 1995). The evolution of the space density
normalisation $\phi^\star$ is harder to model -- the values listed in
Table\ 1 imply little evolution (compared to $L^\star$) with $\left<
  \phi^\star \right> = 1.7\times10^{-3}$\,Mpc$^{-3}$. However, other
surveys have derived a larger range of $\phi^\star$ (e.g.\ Sobral et
al.\ 2009), probably in part due to cosmic variance effects, and the
inherent degeneracy in LF parameter fitting. The latter is the main
reason we chose surveys with very similar fitting techniques; an
attempt to mitigate the impact of different survey strategies on our
model.

With this in mind, the model presented here assumes evolution {\it
  only} in $L^\star$, and the faint end slope is held fixed at
$\alpha=-1.35$ (we assess the impact of this assumption in
\S2.2.1). Given the strong luminosity evolution out to at least
$z=1.3$, and weak evolution beyond to $z\sim2$, we model the $L^\star$
evolution as $(1+z)^Q$ over $0 < z < 1.3$ (the median redshift of the
{\it HST}/NICMOS grism survey of McCarthy et al.\ [1999]). At $z>1.3$
we freeze evolution, and assume this is valid out to the limit of
current H$\alpha$ observations ($z=2.23$). The best fit $L^\star$
evolution is then derived as:
\begin{equation}
  L^\star(z) / {\rm erg\ s^{-1}}=
  \left\{
  \begin{array}{ll}
      
    5.1\times10^{41}\times(1+z)^{3.1\pm 0.4} & z < 1.3 \\
    (6.8^{+2.7}_{-1.9})\times10^{42} & 1.3 < z < 2.2 \\
  \end{array}
  \right.
\end{equation}
We estimate the uncertainty in $Q$ via a bootstrap-type simulation;
re-evaluating the fit 10,000 times after re-sampling each $L^\star$ in
Table\ 1 from a Gaussian distribution of widths set by the $L^\star$
1$\sigma$ uncertainty. Note that $L^\star$ has not been corrected for
intrinsic dust extinction (a canonical $A_{\rm H\alpha}=1$\,mag
correction is generally applied when deriving star formation rates,
although this could increase at high luminosity). The luminosities
{\it have} been corrected for [N~{\sc ii}] contribution, typically of
order $\sim$30\% (e.g.\ Kennicutt \& Kent\ 1983). Note that this could
be a conservative correction if there is a significant contamination
from active galactic nuclei (AGN).  With this in mind, H$\alpha$
redshift surveys should aim for a spectral resolution that can resolve
H$\alpha$/[N~{\sc ii}]. Not only does this have a significant
practical benefit, in that it aids redshift identification, but also
the secondary science impact of a large sample of H$\alpha$/[N~{\sc
  ii}] ratios, and thus AGN selection would be extremely valuable.

As described above, the choice of normalisation of the model is a
source of uncertainty in the predicted counts. Since this paper is
focused on predictions for dark energy surveys, which will target
H$\alpha$ emitters at $z\sim1$, here we have taken the normalisation
of the model to be the average $\phi^\star$ of the surveys of Yan et
al.\ (1999), Hopkins et al.\ (2000) and Shim et al.\ (2009). These three
H$\alpha$ surveys are most similar to the likely observing mode of a
JDEM/Euclid-like mission (slitless spectroscopy), and operate over a
similar redshift range that will be pertinent to cosmology
surveys. The adopted normalisation is
$\phi^\star=1.37\times10^{-3}$\,Mpc$^{-3}$, and in Figure\ 1 we show
how this compares to a range of observed LFs spanning the full
redshift range $0<z<2$. Down to the luminosity corresponding to the
flux limit likely to be practical in cosmology surveys
($\sim$$10^{-16}$\,erg\,s$^{-1}$\,cm$^{-2}$ ), the simple model can
replicate the observed space density of H$\alpha$ emitters over 8\,Gyr
of cosmic time. At fainter limits, the uncertainty in the steepness of
the faint-end slope will introduce further uncertainties that we
ignore here, although we consider the effect of an evolving
(steepening) faint-end in \S2.2.1. 

Figure\ 2 shows another comparison to data that is more relevant for
predictions for dark energy surveys -- i.e. the redshift integrated
counts as a function of limiting flux over $0.75<z<1.90$ (i.e.\
accessible in the near-IR). We compare the integrated counts derived
from the Yan et al.\ (1999), Hopkins et al.\ (2000) and Shim et al.\
(2009) luminosity functions and the model.  Note however, that these
slitless surveys cover much smaller ($<$1\,$\sq^\circ$) areas than
will be achievable with dedicated survey telescopes, and so suffer
significantly from cosmic variance scatter -- this could account for
the scatter in the observations, and highlights the problem of the
choice of normalisation mentioned above. The error-band in our model
does not include the systematic uncertainty due to choice of
$\phi^\star$, but coincidentally spans the range of counts derived
from the surveys shown in Figure\ 2.

For convenience, we tabulate the predicted redshift distribution ${\rm
  d}N/{\rm d}z$ for a range of limiting fluxes, including
uncertainties in Table\ 2. The distributions are plotted in Figure\ 3.
Given the large scatter in the measured space density of H$\alpha$
emitters determined from different surveys (see Hopkins et al.\ 2004
for a compilation), our adopted normalisation should be considered the
best estimate `average'. However, when considering the feasibility of
redshift surveys, the reader might want to adopt a more conservative
estimate of the density normalisation. If necessary, the reader can
re-scale the predicted counts given in Table\ 2. We suggest that an
appropriate conservative lower limit to the counts could be taken as
$\phi^\star=1\times10^{-3}$\,Mpc$^{-3}$. In \S3 we discuss how the
range of adopted normalisations affects our assessment of the
feasibility of redshift surveys that aim to make cosmological
measurements, and in the following section, we address further caveats
that the reader should be aware of when applying this model.

\input{table2.tex}

\subsection{Caveats}

\subsubsection{Evolution of the faint end slope}

Our model assumes a non-evolving faint end slope, with $\alpha=-1.35$
determined from local measurements (Gallego et al.\ 1995; Shioya et
al.\ 2008). Both Yan et al.\ (1999) and Geach et al.\ (2008) fix this
value of $\alpha$ in their fits of the luminosity function; the
observations did not probe deep enough to constrain it. However, there
are hints that the relative abundance of galaxies with $L<L^\star$
might increase towards early epochs (Reddy et al.\ 2008), with
$\alpha$ as large as $-1.6$ at $z\sim2$. What would be the
ramifications of a monotonically evolving (steepening) faint end slope
out to $z\sim2$ on our predicted counts? In Figure\ 3 we compare the
redshift distributions for the fixed $\alpha$ model and the same model
with $\alpha(z)$. We ignore the correlation between $L^\star$,
$\phi^\star$ and $\alpha$ for this analysis. At $f_{\rm lim} >
10^{-16}$\,erg\,s$^{-1}$\,cm$^{-2}$, the counts (per redshift
interval) predicted from the fixed $\alpha$ model are never less than
$\sim$85\% of those derived from a steepening $\alpha$ model. At
$f_{\rm lim} > 5\times10^{-16}$\,erg\,s$^{-1}$\,cm$^{-2}$ the counts
differ by only $\sim$5\%. This difference is smaller than the
uncertainty on ${\rm d}N/{\rm d}z$, and so small enough to be ignored
in this study. Needless to say, as high-$z$ H$\alpha$ studies probe
deeper, past $L^\star$ and can improve the constraint on $\alpha(z)$,
the simple empirical model presented here could be revised
accordingly.  Finally, note that Hopkins et al.  (2000) derive a
faint-end slope of $\alpha=-1.6$, which accounts for the turn-up in
the integrated counts at $f<10^{-16}$\,erg\,s$^{-1}$\,cm$^{-2}$ (Fig.\
2). The empirical model is not significantly different from the Hopkins
et al.\ (2000) counts at brighter limits; re-enforcing that our
assumption of a constant (local) faint-end slope is a reasonable
baseline in this regime.

\subsubsection{Equivalent width cut}

An important feature of emission line surveys, and particularly
narrowband surveys, is the inclusion of an equivalent width (EW) cut
in the selection. Clearly this is an issue of sensitivity: galaxies
with small EW are harder to detect and obtain reliable redshifts
for. So naturally, dark energy surveys targeting emission-lines are
biased towards galaxies with high equivalent widths, and against
weak-emission lines and/or massive galaxies. In the model presented
here we have assumed a fairly low EW cut, 10\,\AA\ in the rest-frame.
This cut will not significantly affect the predicted counts in the
flux regime of interest. For example, according to the model of Baugh
et al.\ (2005), at a flux limit $f_{\rm lim} =
10^{-16}$\,erg\,s$^{-1}$\,cm$^{-2}$, increasing the rest-frame
equivalent width cut from 10\AA\ to 50\AA\ results in a drop in the
number counts (integrated over $0.75<z<1.90$ as in Fig.\ 2) of
$\sim$2\%; the deficit is negligible at brighter limits. In practice,
redshift surveys will probably enforce an observed-frame cut of
$\sim$100\AA.

Finally we note that the clustering properties of bright H$\alpha$
emitters will be different from that of H$\alpha$ emitters with low EW,
or simply continuum- (e.g.\ $H$-band) selected galaxies. The latter
should be more highly biased tracers of the mass distribution (see
Orsi et al.\ 2009 in prep).

\begin{figure}
\centerline{\includegraphics[width=0.49\textwidth]{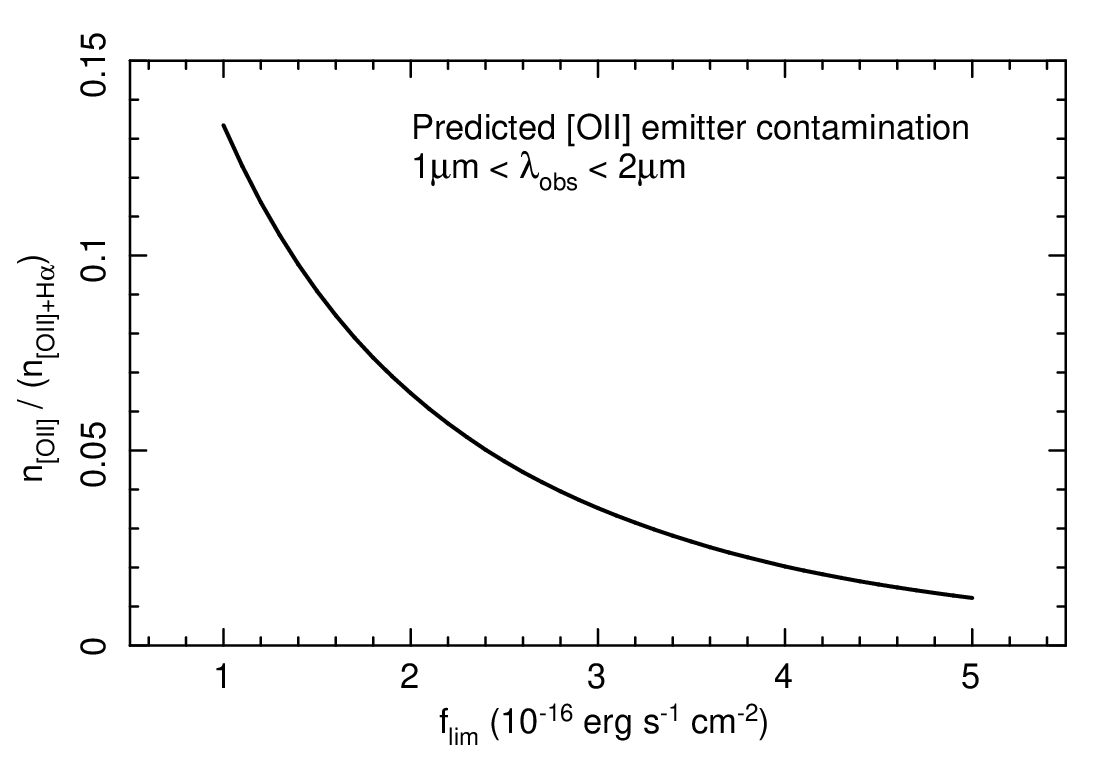}}
\caption{Prediction of contamination from [O~{\sc ii}] emitters over
  the observed wavelength range 1--2$\mu$m. To estimate the number of
  [O~{\sc ii}] emitters, we have adapted the H$\alpha$ count model,
  extrapolating to $z=4.4$ (the redshift of [O~{\sc ii}] at 2$\mu$m)
  making the assumption that all H$\alpha$ emitters are also [O~{\sc
    ii}] emitters, and these galaxies have a constant flux ratio of
  [O~{\sc ii}]/H$\alpha$$=$0.62 (Mouhcine et al.\ 2005). The
  contamination, expressed as a fraction of the total number of
  emitters detected, ranges from 1--13\% in the range of limiting
  fluxes of practical interest.}\end{figure}

\subsubsection{Contamination}

Emission-line surveys (aiming to detect a specific line; in this case
H$\alpha$) are susceptible to contamination from galaxies with {\it
  any} strong emission lines at redshifts placing them in spectral
range of the detector. At high-redshift this can be significantly
problematic -- for example, nearly two thirds of the potential
$z=2.23$ H$\alpha$ emitters of Geach et al.\ (2008) selected with a
narrowband at 2.121$\mu$m were eliminated as low-redshift contaminants
(e.g.\ Pa$\alpha$ [$z=0.13$], Pa$\beta$ [$z=0.67$], Fe{\sc ii}
[$z=0.3$]). Higher redshift [O~{\sc iii}]$\lambda$5007 can also
contribute to the contamination. Geach et al.\ (2008) used further
broad-band colour- and luminosity selections to select the $z=2.23$
candidates. Although most planned dark energy surveys will employ
spectroscopy, one must still consider the potential for
mis-identification of the H$\alpha$ line in the large redshift ranges
these surveys will probe.

One could use the H$\alpha$ model presented here to estimate the
potential level of mis-identification of emission lines in spectral
ranges likely to be employed in a slitless survey. For example,
consider contamination from [O~{\sc ii}] emitters at a rest-frame
wavelength of 3727\AA. For a survey operating at 1--2$\mu$m, this
means contamination from galaxies in the redshift range $1.7 < z <
4.4$. If we assume that every H$\alpha$ emitter is also an [O~{\sc
  ii}] emitter, then we can estimate the expected number of objects in
addition to the H$\alpha$ emitters detected, assuming an attenuation
due to the flux ratio [O~{\sc ii}]/H$\alpha$$<$1 and intrinsic
extinction $A_{\rm [OII]}$. In this example, we assume [O~{\sc
  ii}]/H$\alpha$$=$0.62 (measured from the 2dF Galaxy Redshift survey,
at $z\sim0.06$; Mouhcine et al.\ 2005). Note that this ratio has not
been corrected for the relative intrinsic extinction, and so this
prediction should reflect the actual number of galaxies a flux-limited
survey can expect to detect\footnote{Although strictly our model for
  the abundance of H$\alpha$ emitters only extends to $z\sim2$, we
  assume the fixed evolution extends to $z=4.4$. If the number of
  H$\alpha$ emitters is actually gently declining at $z>2$, this
  contamination estimate should be considered a conservative upper
  limit.}. As a fraction of the total number of emitters detected, the
contamination from [O~{\sc ii}] emitters ranges between 13\% for a
limiting flux of $10^{-16}$\,erg\,s$^{-1}$\,cm$^{-2}$, to $\sim$1\%
for $5\times10^{-16}$\,erg\,s$^{-1}$\,cm$^{-2}$. A plot of the decline
in contamination as a function of limiting flux is shown for reference
in Figure\,4.

There are two simple ways to mitigate contamination. Perhaps the most
efficient way to identify H$\alpha$ is to resolve the [N~{\sc
  ii}]$\lambda$6583 line (offset $\Delta\lambda = 20$\AA\ from
H$\alpha$). Identifying this pair of lines is a useful discriminant
between H$\alpha$ and `contaminant' lines, and so dark energy surveys
should aim for a spectral resolution of $R>500$ to achieve
this. Another aid to redshift determination is the new generation of
all sky ground based photometric surveys (e.g.\ PanSTARRS, Large
Synoptic Survey Telescope). These surveys will provide optical
photometry of many of the sources detected in the dark energy surveys;
in conjunction with the near-IR photometry this will improve redshift
estimates with a photo-$z$ technique.

\subsubsection{Extinction}

The high redshift H$\alpha$ surveys described in this work have not
been corrected for {\it intrinsic} dust extinction, although when
deriving star formation rates, many authors tend to apply a canonical
$A_{\rm H\alpha}=1$\,mag unless some better estimate exists. The
predicted number counts in our simple model include this intrinsic
extinction, such that if the extinction properties of the H$\alpha$
emitters in the surveys described in Table\ 1 are relatively constant
over a wide range of redshift, then the predicted counts can be taken
as a reliable representation of the expected yield even considering
internal extinction. However, all sky surveys (even ones that exclude
the Galactic plane) will encounter a range of foreground Galactic
extinction.  Despite H$\alpha$ being redshifted into the
near-infrared at $z>0.5$, where reddening is fairly negligible, for
completeness we consider here whether this could impact the predicted
counts.

Taking the all-sky dust maps of Schlegel, Finkbeiner \& Davis\
(1998)\footnote{\tt irsa.ipac.caltech.edu/applications/DUST/}, we
evaluate the $V$-band extinction for Galactic latitudes
$|b|>20$$^\circ$, and extrapolate this to the observed wavelength of
H$\alpha$, $\lambda = (1+z)\times 6563$\AA\ out to $z=2.2$ assuming a
$R_V=3.1$ reddening law for the Galaxy (Cardelli et al.\ 1989;
O'Donnell\ 1994). For reference, we summarise the average and range of
reddenings for each redshift bin in Table\ 2. Of course, at longer
wavelengths (in other words, H$\alpha$ observed at higher-redshifts)
reddening has an ever decreasing impact on the effective flux limit:
at $z>0.5$ the maximum $A_{\rm H\alpha}$ is never more than 0.2\,mag,
and the average is always $<$0.03\,mag.

Since the regions of `high' reddening represent a small fraction of
the extragalactic sky, Galactic reddening has a minor (though redshift
dependent) impact on the predicted counts. For example, modelling the
variation in $A_{\rm H\alpha}$ over the full $|b|>20^\circ$ sky, at
$z=0.5$ there is only a 2\% decline in ${\rm d}N/{\rm d}z$; a smaller
variation than the uncertainty of our model -- we ignore its effects.

\begin{figure}
\centerline{\includegraphics[width=0.49\textwidth]{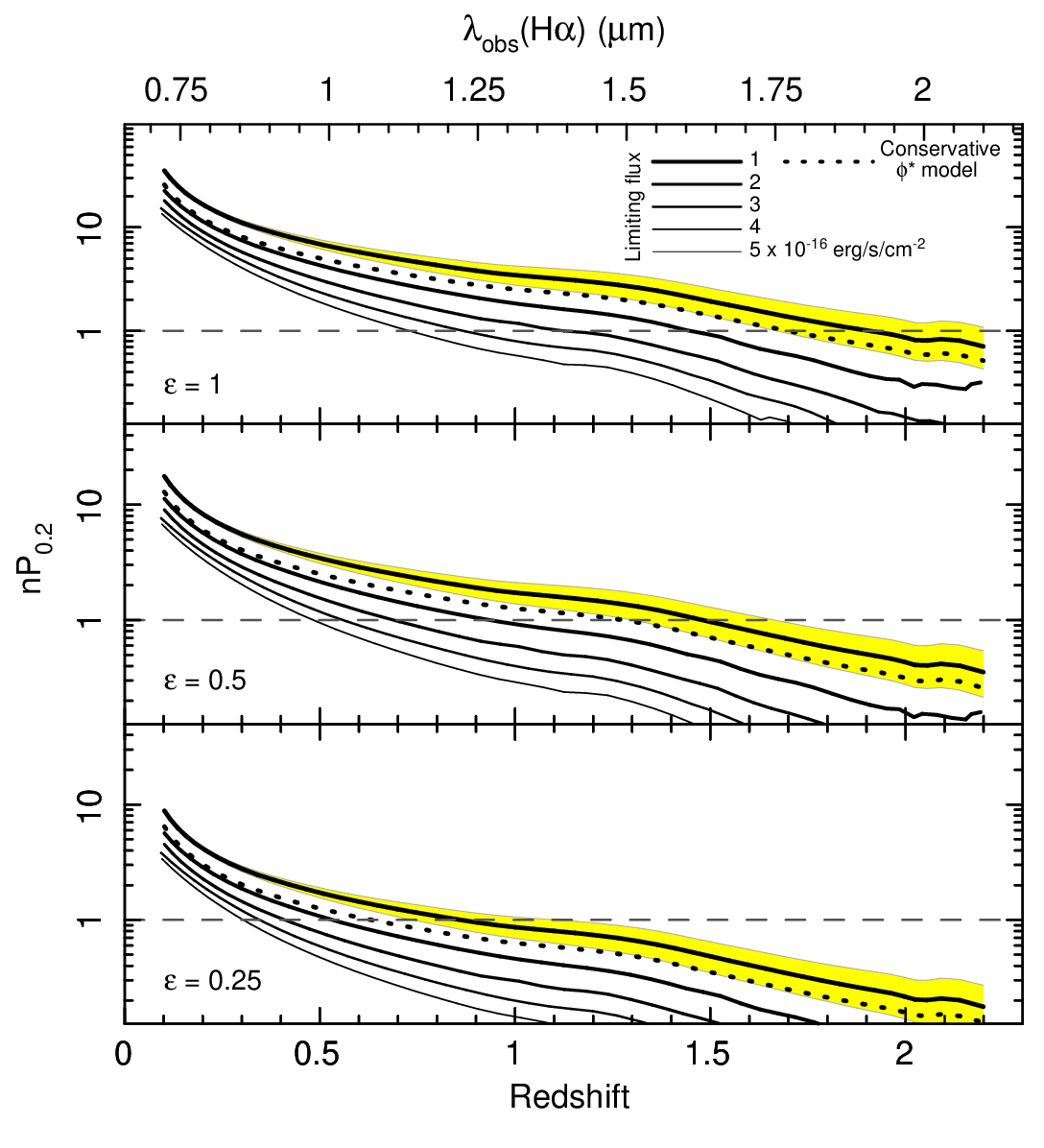}}
\caption{Predictions for the effective power of a galaxy redshift
  survey, expressed in terms of the shot-noise parameter $\bar{n}P$
    evaluated at $k=0.2$\,$h$\,Mpc$^{-1}$ (approximately the peak of
    the BAO signal). Fixed-time redshift surveys should aim for the
    sweet-spot of $\bar{n}P_{0.2}=1$ to obtain maximum power from the
    survey. We show the predicted $\bar{n}P_{0.2}$ for limiting fluxes
    of $1$--$5$$\times10^{-16}$\,erg\,s$^{-1}$\,cm$^{-2}$, and three
    survey `efficiencies' ($\epsilon$: the actual sampling of the
    H$\alpha$ population due to the success rate of the survey). Note
    the clear degeneracy between survey efficiency and flux limit. The
    solid lines show the predictions for our `average' model
    $\phi^\star$ normalisation, but we also show the predicted
    $\bar{n}P_{0.2}$ for a more conservative normalisation,
    $\phi^\star=10^{-3}$\,Mpc$^{-3}$ (for clarity only shown for
    $f_{\rm lim}=10^{-16}$\,erg\,s$^{-1}$\,cm$^{-2}$).  The conclusion
    to draw from this plot is that H$\alpha$ surveys should be aiming
    for flux limits of $\sim$$10^{-16}$\,erg\,s$^{-1}$\,cm$^{-2}$;
    beyond $z\sim1$ the redshift yield goes into sharp decline, with
    severe consequences for $\bar{n}P_{0.2}$.} \end{figure}

\section{Implications for redshift surveys}

Dark energy surveys that aim to detect BAOs and measure redshift
distortions in galaxy clustering could target H$\alpha$ emitters in
all-sky near-infrared surveys, most likely utilising grisms for
slitless spectroscopy (e.g.\ McCarthy et al.\ 1999, Fig.\ 2). The key
issue for these surveys is the ability to measure sufficient numbers
of redshifts for an accurate assessment of $w(z)$ and $f(z)$. Let us
consider a hypothetical example: a slitless survey from a space
platform with a wavelength coverage of 1--2$\mu$m, and a spectral
resolution of $R>500$. This range gives access to H$\alpha$ at $0.5 <
z < 2$, with sufficient resolution to resolve [N~{\sc
  ii}]$\lambda$6583 at $f_{\rm lim} >
10^{-16}$\,erg\,s$^{-1}$\,cm$^{-2}$. Aside from the slight
modification to nominal limiting flux due to Galactic extinction
(\S2.2.4), there should be an additional modification to predicted
counts due to some non-unity efficiency factor $\epsilon$ (the ratio
of the number of successfully measured redshifts, to the total number
of measurable redshifts at a given flux limit). This will inevitably
vary as a function of flux, equivalent width, and so on).  Including
some assumption for $\epsilon$, how optimistic can we be about
measurements of $w(z)$ and $f(z)$ in redshift surveys?

A precise measurement of $w(z)$ or $f(z)$ requires an accurate
measurement of the power spectrum, $P(k)$. The uncertainty with which
$P(k)$ can be measured from a given galaxy survey depends on the
number density of galaxies and the volume of the survey. If the number
density is low, then the errors are dominated by shot noise. If it is
high, then cosmic variance (i.e.\ the volume of the survey) dominates
the error budget. To see this, note that the effective volume of a
survey is given by Feldman, Kaiser \& Peacock\ (1994)
as: \begin{equation} V_{\rm eff} = \int {\rm
    d}^3r\left[\frac{\bar{n}({\bf r})\bar{P}} {1+\bar{n}({\bf
        r})\bar{P}}\right]^2,
\end{equation}
where $\bar{n}({\bf r})$ is the comoving number density of the sample
at location ${\bf r}$. For small $\bar{n}$, $V_{\rm
  eff}\propto\bar{n}$ and the signal is shot noise dominated. For
large $\bar{n}$, $V_{\rm eff}=V$, where $V$ is the physical volume of
the survey, which limits the signal. For a sample with a fixed total
number of galaxies $N_{\rm gal}=\bar{n}V$, and for a power spectrum
${P}$, setting ${\rm d}V_{\rm eff}/{\rm d}V=0$ requires
$\bar{n}P=1$. In this situation we see that the effective volume
reaches a maximum when $\bar{n}P=1$. This `sweet-spot' is often used
as a design aim for fixed integration-time and/or volume limited
galaxy redshift surveys, with ${P_{0.2}}\equiv\langle
P(k)\rangle$, calculated for $k=0.2\,h$\,Mpc$^{-1}$. This scale is
approximately the limit of the quasi-linear regime, and this also
gives an indication of the strength of the clustering signal on the
linear scales carrying the redshift-distortion information.

Future surveys will often be limited by the extragalactic sky area
they can observe. For surveys using a single ground-based telescope
(such as BOSS) this is of order $\sim$$10^4$ square degrees, while for
a space based platform (such as Euclid or JDEM) or a survey using a
pair of telescope in different hemispheres, this is
$\sim$$(2-3)\times10^4$ square degrees. In this situation, the volume
that can be surveyed in the interesting redshift range is limited, and
the only way of gaining signal is to push to higher galaxy number
densities. It is therefore important to consider values of
$\bar{n}P_{0.2}>1$.

In Figure\ 5 we show the predicted $\bar{n}P_{0.2}$ as a function of
redshift, for a range of (nominal) limiting fluxes
$(1$--$5)\times10^{-16}$\,erg\,s$^{-1}$\,cm$^{-2}$. As well as the
ideal case, with an efficiency factor $\epsilon=1$ (that is, one
correctly identifies all the H$\alpha$ emitters above the survey flux
limit in every pointing), we show the effect on $\bar{n}P_{0.2}$ for a
50\% and 25\% efficiency. Note that we have assumed a model for the
luminosity-dependent evolution of bias for H$\alpha$ emitters from
Orsi et al.\ (2009) such that $P_{\rm gal} = P_{\rm DM}b(z,L_{\rm
  H\alpha})^2$. The H$\alpha$ population is generated by the
semi-analytic prescription {\tt GALFORM} (Baugh et al.\ 2005). Since
in the semi-analytic model one can ask what dark matter halo hosts a
given galaxy, Orsi et al.\ estimate the galaxy bias for a given
H$\alpha$ luminosity by averaging over the halos that host selected
H$\alpha$ emitters. The model bias for H$\alpha$ emitters at $z\sim2$
agrees well with the value derived by Geach et al.\ (2008) from the
projected two-point correlation function. Note that we have applied
the same rest-frame EW cut as applied throughout this work, and
interpolated the $b(z,L_{\rm H\alpha})$ as necessary. As a guide, the
range of bias applied over $0 < z < 2$ for the luminosities
corresponding to the limiting fluxes considered here is $0.9\lesssim
b\lesssim 1.7$.

Obviously one would always strive for maximum efficiency and depth,
but this is not a practical possibility: there will always be redshift
attrition resulting in $\epsilon<1$. This inefficiency has the same
impact as increasing the effective limiting flux of the survey. Losing
counts has a serious impact on the survey power; even at the faintest
limit likely to be practicable, $10^{-16}$\,erg\,s$^{-1}$\,cm$^{-2}$,
a `perfect' survey struggles to achieve $\bar{n}P_{0.2}=1$ at
$z=2$. Assuming the more likely case of $\epsilon=0.5$, one can
comfortably achieve the required $\bar{n}P_{0.2}$ out to $z=1$, even
with fairly conservative flux limits. At higher redshifts this becomes
increasingly observationally expensive. Re-visiting the caveat of
model normalisation described in \S2.1, on Figure\ 5 we also show the
more conservative case the reader might choose to adopt. Obviously a
shift in normalisation simply translates the predicted
$\bar{n}P_{0.2}$ up or down. It is worth noting that the conservative
counts are within the 1$\sigma$ band of uncertainty of the average
model normalisation at $z\geq1$, and so our conclusions about the
power of redshift surveys as a function of limiting flux and
efficiency are unchanged.

One way to boost performance would be to employ Digital Micro-mirror
Devices (DMDs), rather than traditional slitless spectroscopy. For a
fixed telescope diameter and integration time, with DMD-slit
spectroscopy one reaches $\sim$2.5\,mag deeper in the continuum, due
to the strong reduction of the sky background compared to slitless
spectroscopy. This allows the detection of several spectral features
in each spectrum (absorption and emission lines) and the consequent
identification of all galaxy types (early-type and star-forming
systems). Moreover, thanks to improved sensitivity and the lack of the
`spectral confusion' problem due to the overlap of spectra of
different objects (the traditional Achille's heel of slitless
spectroscopy), the redshift success rate $\epsilon$ is much higher (up
to $>$90\%, see Cimatti et al. 2009).

\section{Cosmological near-IR surveys from the ground}

To be competitive with space-platforms targeting H$\alpha$ emitters at
$z>0.5$, ground-based near-IR BAO surveys should also be aiming for
limiting fluxes of $10^{-16}$\,erg\,s$^{-1}$\,cm$^{-2}$, but there are
extra observational challenges -- not least the deleterious effect of
the atmosphere in the near-IR. Approximately 30\% of the 1--2$\mu$m
window has an atmospheric transmission of $<$80\%, mainly affecting
H$\alpha$ in the redshift ranges $1<z<1.3$ and $1.7<z<2.1$. In
addition, near-IR observations from the ground must also contend with
forest of OH-airglow: even at $R\sim2000$ less than half of the
near-IR spectral range is free from OH line emission, although new OH
suppression technologies could partly mitigate this effect.

On the basis of areal coverage, ground-based near-IR BAO survey will
never be competitive with a Euclid/JDEM-like mission. Modern
wide-field near-IR spectrographs deploy fibres on individual targets,
and this presents a significant disadvantage compared to the slitless
approach of the space missions: one must select targets prior to
observation (in some sense the problem is reversed in the slitless
case). Typically this will require the target fields to be
complemented by multi-colour broad-band photometry, deep enough to
provide an estimate of redshift. Note that the consequence for
mis-identifying targets is a strong hit to the efficiency parameter
$\epsilon$.

In the event of a dedicated space-based near-infrared dark energy
survey going ahead, one could argue that a more efficient use of
ground based multi-object spectrographs in the near-IR would be to
complement the wider cosmological surveys by providing more detailed
follow-up observations of a sub-sample of line-emitters. This has the
advantage of side-stepping the issue of target selection, since the
sample would already be `sanitized' by the cosmology survey. Such a
symbiosis between space and ground would be an efficient use of
resources since: (a) the ground facilities would target known
line-emitters, and therefore rapidly build up a large sample of
spectroscopic observations for high-redshift galaxies in more detail
than can be achieved from the space platforms; and (b) the
complementary observations could help to better characterise
contamination from other line emitters (as discussed in \S2.2.3), thus
feeding back information to the cosmological survey. Combining surveys
in this way could serve to satisfy two groups of researchers: those
interested in the astrophysics of galaxies at high redshift, and those
concerned with cosmological measurements.

\section{Summary and final remarks}

We have presented a simple prescription for the prediction of the
abundance of H$\alpha$ emitters over $0 < z < 2$, based on empirical
data. The model is simplistic, due to limited available data; it
assumes a fixed space density, fixed faint end slope, and only
$L^\star$ evolution out to $z=1.3$. There is no luminosity evolution
to higher redshifts, consistent with current H$\alpha$ observations at
this redshift. Despite its simplicity, the model adequately mimics the
observed luminosity functions of a range of H$\alpha$ surveys
(including a mixture of spectroscopic, grism and narrowband
strategies). Using the luminosity function model as a basis, we
predict the redshift distribution of H$\alpha$ emitters corresponding
to a spectral coverage that extends to 2$\mu$m.

Our results have particular relevance to dark energy experiments
attempting to measure cosmological information from the power spectrum
of galaxies detected in all-sky H$\alpha$ surveys in the near-IR. We
use the parameter $\bar{n}P_{0.2}$ as a measure of the effectiveness
of a redshift survey, and make predictions for this value for a range
of redshift, limiting flux and success rate (i.e.\ efficiency). To
achieve $\bar{n}P_{0.2}=1$ out to $z=2$, emission-line surveys should
be aiming for limiting fluxes of
$\sim$10$^{-16}$\,erg\,s$^{-1}$\,cm$^{-2}$. However, this estimate is
reliant on a high success rate of the sampling of the H$\alpha$
population: redshift surveys need to aim for high-efficiencies, since
any decline in redshift yield (i.e.\ failing to obtain redshifts for
detections) has the same effect on $\bar{n}P_{0.2}$ as increasing the
flux limit (illustrated in Figure\ 5 of this work). Assuming a more
likely situation of 50\% efficiency, a realistic target for proposed
surveys is $\bar{n}P_{0.2}=1$ at $z=1.5$. At higher redshifts the
sharply declining number counts have a severe effect on one's ability
to measure $w(z)$ at the desired precision.

\section*{Acknowledgements}

We thank the referee for useful comments. The authors would also like
to thank H. Shim, Lin\ Yan \& Philip\ Hopkins for helpful
discussions. JEG is funded by the U.K. Science and Technology
Facilities Council (STFC). WJP is grateful for support from the STFC,
the Leverhulme Trust and the European Research Council. AC, BG, GZ,
LG, LP, and PF acknowledge the support from the Agenzia Spaziale
Italiana (ASI, contract N. I/058/08/0)

\label{lastpage}

\end{document}

%% file: table2.tex
\begin{table*}
  \caption{Redshift distributions ${\rm d}N/{\rm d}z$ (per square degree,         calculated
    in bins of width $\delta z = 0.1$, centred on the value given in the first         column) for a
    range of limiting fluxes derived from the empirical model (also see         Figure\ 3). For reference,     we provide the range of Galactic extinctions at         the observed wavelength                     of H$\alpha$, derived from     the         maps of Schlegel, Finkbeiner \& Davis (1998). The predicted counts include         {\it intrinsic}                                         extinction                             in the H$\alpha$ emitters, but the Galactic reddening will vary as a function         of sky position. Although this has a negligible (few per cent) impact on the         model         ${\rm                   d}N/{\rm d}z$, we include it here as a         guide. The counts listed here are calculated for a space density normalisation         of         $\phi^\star=1.37\times10^{-3}$\,Mpc$^{-3}$             which is the         `average' space density of H$\alpha$ emitters determined by several slitless         surveys at $z\sim1$ -- similar to     the     Euclid and JDEM satellite survey                     concepts. The reader can re-scale these counts to alternative normalisations         if     desired: for a more conservative estimate of the counts, we recommend a         lower density         normalisation $\phi^\star=1\times10^{-3}$\,Mpc$^{-3}$,         however as we show in \S3, this choice does not have a significant impact on         the predicted power of a galaxy redshift     survey.
  }
\begin{center}
\begin{tabular}{@{\extracolsep{\fill}}crrrrrccc}
\hline
& \multicolumn{5}{c}{Number per $\delta z= 0.1$ interval
  (deg$^{-2}$)}& \multicolumn{3}{c}{Reddening at $(1+z)\times6563$\AA}\cr

& \multicolumn{5}{c}{Limiting flux ($\times$10$^{-16}$\
  erg\,s$^{-1}$\,cm$^{-2}$)}& \multicolumn{3}{c}{$|b|>20^\circ$ (mag)}\cr
Redshift & \multicolumn{1}{c}{1} & \multicolumn{1}{c}{2}
&\multicolumn{1}{c}{3}  &\multicolumn{1}{c}{4}  &
\multicolumn{1}{c}{5}& $A^{\rm min}_{\rm H\alpha}$ & $\left<A_{\rm H\alpha}\right>$ & $A^{\rm max}_{\rm H\alpha}$ \cr
\hline
0.10 & $5226^{+86}_{-85}$ & $3838^{+67}_{-66}$ & $3172^{+58}_{-57}$ & $2756^{+52}_{-52}$ & $2461^{+48}_{-47}$ & 0.005 & 0.045 & 0.293\cr
0.20 & $10160^{+367}_{-357}$ & $7116^{+284}_{-277}$ & $5669^{+244}_{-237}$ & $4771^{+218}_{-211}$ & $4142^{+199}_{-193}$ & 0.005 & 0.040 & 0.260\cr
0.30 & $14448^{+834}_{-801}$ & $9702^{+639}_{-612}$ & $7473^{+541}_{-517}$ & $6107^{+478}_{-456}$ & $5163^{+431}_{-410}$ & 0.004 & 0.036 & 0.231\cr
0.40 & $17931^{+1446}_{-1372}$ & $11592^{+1094}_{-1033}$ & $8657^{+915}_{-860}$ & $6885^{+798}_{-746}$ & $5676^{+712}_{-661}$ & 0.004 & 0.032 & 0.207\cr
0.50 & $20673^{+2160}_{-2023}$ & $12915^{+1613}_{-1498}$ & $9376^{+1332}_{-1227}$ & $7270^{+1147}_{-1047}$ & $5854^{+1010}_{-914}$ & 0.003 & 0.029 & 0.186\cr
0.60 & $22787^{+2936}_{-2714}$ & $13803^{+2165}_{-1976}$ & $9766^{+1766}_{-1592}$ & $7398^{+1502}_{-1336}$ & $5830^{+1306}_{-1147}$ & 0.003 & 0.026 & 0.168\cr
0.70 & $24386^{+3741}_{-3414}$ & $14368^{+2727}_{-2446}$ & $9931^{+2199}_{-1938}$ & $7365^{+1847}_{-1600}$ & $5689^{+1587}_{-1351}$ & 0.003 & 0.023 & 0.152\cr
0.80 & $25574^{+4553}_{-4100}$ & $14699^{+3283}_{-2891}$ & $9947^{+2618}_{-2254}$ & $7236^{+2175}_{-1832}$ & $5489^{+1849}_{-1523}$ & 0.002 & 0.021 & 0.138\cr
0.90 & $26437^{+5353}_{-4759}$ & $14861^{+3821}_{-3305}$ & $9868^{+3017}_{-2538}$ & $7054^{+2482}_{-2032}$ & $5263^{+2089}_{-1663}$ & 0.002 & 0.019 & 0.125\cr
1.00 & $27045^{+6128}_{-5381}$ & $14905^{+4335}_{-3683}$ & $9730^{+3392}_{-2788}$ & $6847^{+2766}_{-2200}$ & $5034^{+2308}_{-1776}$ & 0.002 & 0.018 & 0.114\cr
1.10 & $27456^{+6872}_{-5960}$ & $14867^{+4823}_{-4025}$ & $9558^{+3744}_{-3007}$ & $6632^{+3029}_{-2341}$ & $4811^{+2506}_{-1865}$ & 0.002 & 0.016 & 0.104\cr
1.20 & $27712^{+7579}_{-6495}$ & $14772^{+5282}_{-4332}$ & $9369^{+4071}_{-3196}$ & $6420^{+3270}_{-2458}$ & $4601^{+2687}_{-1933}$ & 0.002 & 0.015 & 0.096\cr
1.30 & $27850^{+8248}_{-6986}$ & $14641^{+5712}_{-4606}$ & $9173^{+4375}_{-3359}$ & $6216^{+3493}_{-2554}$ & $4407^{+2853}_{-1986}$ & 0.002 & 0.014 & 0.088\cr
1.40 & $24931^{+7883}_{-6612}$ & $12514^{+5325}_{-4210}$ & $7530^{+3978}_{-2965}$ & $4913^{+3098}_{-2177}$ & $3360^{+2468}_{-1636}$ & 0.001 & 0.012 & 0.081\cr
1.50 & $22178^{+7487}_{-6211}$ & $10600^{+4919}_{-3805}$ & $6108^{+3574}_{-2579}$ & $3827^{+2708}_{-1822}$ & $2517^{+2099}_{-1318}$ & 0.001 & 0.011 & 0.074\cr
1.60 & $19621^{+7071}_{-5796}$ & $8905^{+4505}_{-3402}$ & $4900^{+3176}_{-2210}$ & $2939^{+2334}_{-1497}$ & $1854^{+1755}_{-1038}$ & 0.001 & 0.011 & 0.068\cr
1.70 & $17272^{+6645}_{-5374}$ & $7422^{+4095}_{-3011}$ & $3888^{+2792}_{-1868}$ & $2226^{+1985}_{-1209}$ & $1343^{+1444}_{-801}$ & 0.001 & 0.010 & 0.063\cr
1.80 & $15136^{+6215}_{-4955}$ & $6141^{+3693}_{-2640}$ & $3053^{+2429}_{-1557}$ & $1664^{+1666}_{-958}$ & $956^{+1169}_{-604}$ & 0.001 & 0.009 & 0.059\cr
1.90 & $13209^{+5788}_{-4543}$ & $5044^{+3307}_{-2292}$ & $2373^{+2091}_{-1280}$ & $1227^{+1380}_{-747}$ & $670^{+932}_{-446}$ & 0.001 & 0.008 & 0.054\cr
2.00 & $11482^{+5368}_{-4144}$ & $4114^{+2940}_{-1971}$ & $1826^{+1783}_{-1039}$ & $892^{+1128}_{-572}$ & $461^{+731}_{-323}$ & 0.001 & 0.008 & 0.050\cr
2.10 & $9945^{+4960}_{-3761}$ & $3332^{+2596}_{-1680}$ & $1390^{+1505}_{-832}$ & $640^{+911}_{-430}$ & $312^{+564}_{-228}$ & 0.001 & 0.007 & 0.047\cr
2.20 & $8582^{+4566}_{-3397}$ & $2681^{+2277}_{-1418}$ & $1048^{+1258}_{-657}$ & $453^{+726}_{-318}$ & $208^{+429}_{-158}$ & 0.001 & 0.007 & 0.043\cr\hline
\end{tabular}
\end{center}
\end{table*}

%% file: geach_euclid_astroph.bbl
\begin{thebibliography}{10}


\bibitem[\protect\citeauthoryear{Albrecht et al.}{2006}]{albrecht06}
    Albrecht A., et al., 2006, {\it Report of the Dark Energy Task
    Force}, astro-ph/0609591

\bibitem[]{}
{Baugh, C. M., Lacey, C. G., Frenk, C. S., Granato, G. L., Silva, L.,
  Bressan, A., Benson, A. J., Cole, S., 2005, \mnras, 356, 1191}

\bibitem[]{} {Blake, C. \& Glazebrook, K., 2003, \apj, 594, 665}

 
  \bibitem[\protect\citeauthoryear{Bond \& Efstathiou}{1984}]{bond84}
    Bond, J.R. \& Efstathiou, G. 1984, ApJ, 285, L45
  \bibitem[\protect\citeauthoryear{Bond \& Efstathiou}{1987}]{bond87}
    Bond, J.R., \& Efstathiou, G., 1987, MNRAS, 226, 655

\bibitem[]{}
{Bower, R. G., Benson, A. J., Malbon, R., Helly, J. C., Frenk, C. S.,
  Baugh, C. M., Cole, S., Lacey, C. G., 2006, \mnras, 370, 645}

\bibitem[]{}{Cardelli, J.~A., Clayton, G.~C., Mathis, J.~S., 1989,
    \apj, 345, 245}

 
  \bibitem[\protect\citeauthoryear{Cole et al.}{2005}]{cole05}
    Cole S., et al., 2005, MNRAS, 362, 505
  \bibitem[\protect\citeauthoryear{Colless et al.}{2003}]{colless03}
    Colless M., et al., 2003, astro-ph/0306581

\bibitem[]{} {Dalton, G. B., et. al., 2006, Ground-based and Airborne
    Instrumentation for Astronomy. Eds. McLean, I.~S., Iye,
    M. Proceedings of the SPIE, Volume 6269, 62694A}

\bibitem[]{}{Dvali, G.\ R., Gabadadze, G., Porrati, M., 2000, Phys.\ Lett.\ B, 485,
 208 }

\bibitem[]{} {Eisenstein, D.~J., Hu, W., 1998, \apj, 496, 605}

\bibitem[]{}
{Eisenstein, D., 2002, 
	Next Generation Wide-Field Multi-Object Spectroscopy, ASP
        Conference Proceedings, Eds. M. J. I. Brown \& A.
        Dey. Astronomical Society of the Pacific, 280, 35}


  \bibitem[\protect\citeauthoryear{Eisenstein et al.}{2005}]{eisenstein05}
    Eisenstein D.J., et al., 2005, ApJ, 633, 560
  \bibitem[\protect\citeauthoryear{Eisenstein et al.}{2007}]{eisenstein07}
    Eisenstein D.J., Seo H.-J., White M., 2007, ApJ, 664, 660

    \bibitem[]{} Eto, S., et al., 2004, Proc. SPIE, 5492, 1314

\bibitem[]{}
{Feldman, H.~A., Kaiser, N., Peacock, J.~A., 1994, \apj, 426, 23}

\bibitem[]{}
{Gallego}, J., {Zamorano}, J., {Aragon-Salamanca}, A., 
	{Rego}, M., 1995, \apjl, 455, L1

  \bibitem[\protect\citeauthoryear{Gaztanaga et al.}{2008}]{gaztanaga08}
    Gaztanaga E., Cabre A., Hui L., 2009, \mnras, 399, 1663

\bibitem[]{}
{Geach, J. E., Smail, Ian, Best, P. N., Kurk, J., Casali, M., Ivison,
  R. J., Coppin, K., 2008, \mnras, 388, 1473}


\bibitem[\protect\citeauthoryear{Glazebrook et
    al.}{2007}]{glazebrook07} Glazebrook K., et al., 2007, ASP
  conference series, 379, 72


  \bibitem[\protect\citeauthoryear{Goldberg \& Strauss}{1998}]{goldberg98}
    Goldberg D.M., Strauss M.A., 1998, ApJ, 495, 29

  \bibitem[]{} {Guzzo, L. et al., 2008, \nat, 451, 541}

  \bibitem[\protect\citeauthoryear{Hicken et al.}{2009}]{hicken09} 
    Hicken M., et al., 2009, ApJ, 700, 1097
  \bibitem[\protect\citeauthoryear{Hill et al.}{2008}]{hill08} Hill
    G.~J., et al., 2008, ASP conference series, 399, 115
   


  \bibitem[\protect\citeauthoryear{Holtzman}{1989}]{holtzman89}
    Holtzman J.A. 1989, ApJS, 71,1

\bibitem[]{} {Hopkins, A.~M., Connolly, A.~J., Szalay, A.~S.,
          2000, \aj, 120, 2843}

\bibitem[]{}
{Hopkins}, A.~M., 2004, \apj, 615, 209


  \bibitem[\protect\citeauthoryear{Hu \& Haiman}{2006}]{hu03} 
    Hu W., Haiman Z., 2003, PRD, 68, 3004
  \bibitem[\protect\citeauthoryear{Huetsi}{2006}]{huetsi06} 
    Huetsi G., 2006, A\&A, 449, 891

\bibitem[]{}
{Kennicutt}, Jr., R.~C., {Kent}, S.~M., 1983, \aj, 88, 1094

\bibitem[]{}
{Lilly}, S.~J., {Tresse}, L., {Hammer}, F., {Crampton}, D., 
	{Le Fevre}, O., 1995, \apj, 455, 108

\bibitem[]{}Ly, C., Malkan, M. A., Kashikawa, N., Shimasaku,
  K., Doi, M., Nagao, T., Iye, M., Kodama,
  T., Morokuma, T., Motohara, K., 2007, \apj, 657, 738


\bibitem[]{}
{McCarthy, P., et al., 1999, \apj, 520, 548}


  \bibitem[\protect\citeauthoryear{Meiksin et al.}{1999}]{meiksin99}
    Meiksin A., White M. \& Peacock J.A., 1999, MNRAS, 304, 851

\bibitem[]{}
{Moorwood}, A.~F.~M., {van der Werf}, P.~P., {Cuby}, J.~G., 
	{Oliva}, E., 2000, \aap, 326, 9

      \bibitem[]{} {O'Donnell, J.~E., 1994, \apj, 422, 1580}


 \bibitem[]{} {Orsi, A., et al.\ 2009, \mnras, submitted}


\bibitem[]{}
{Peebles, P. J., Ratra, B., 2003, Reviews of Modern Physics, 75, 559}

\bibitem[]{}
{Peebles, P. J., Yu, J., T., 1970, \apj, 162, 815}


  \bibitem[\protect\citeauthoryear{Percival et al.}{2001}]{percival01}
    Percival, W.~J., et al., 2001, MNRAS, 327, 1297
  \bibitem[\protect\citeauthoryear{Percival et al.}{2007}]{percival07} 
    Percival, W.~J., Cole S., Eisenstein D., Nichol R., Peacock J.A.,
    Pope A., Szalay A., 2007, MNRAS, 381, 1053
  \bibitem[\protect\citeauthoryear{Percival et al.}{2009}]{percival09}
    Percival, W.~J., et al., 2009, MNRAS~submitted,\
    arXiv:0907.1660


\bibitem[]{}
{Perlmutter, S., et al.\ 1999, \apj, 517, 565}

\bibitem[]{}
{Reddy}, N.~A., et al., 2008, \apj, 175, 48.

\bibitem[]{}
{Riess, A.~G., et al.\ 1998, \aj, 116, 1009}

\bibitem[]{}
{Seo, H.-J., Eisenstein, D. J., 2003, \apj 498, 720}

\bibitem[\protect\citeauthoryear{Seo \& Eisenstein}{2005}]{seo05} Seo,
  H.-J., Eisenstein D.J., 2005, ApJ, 633, 575

\bibitem[]{}
{Schlegel, D.~J., Finkbeiner, D.~P., Davis, M., 1998, \apj, 500, 525}


  \bibitem[\protect\citeauthoryear{Schlegel et al.}{2009}]{schlegel09}
    Schlegel D., White M., Eisenstein D.J., 2009, [[arXiv:0902.4680]]


\bibitem[]{}
{Shim, H., Colbert, J., Teplitz, H., Henry, A., Malkan,
  M., McCarthy, P., Yan, L., 2009, \apj, 696, 785}

\bibitem[]{}
{Shioya, Y., et al., 2008, \apjs, 175, 128}


  \bibitem[\protect\citeauthoryear{Silk}{1968}]{silk68}
    Silk J., 1968, ApJ, 151, 459


  \bibitem[\protect\citeauthoryear{Springel et al.}{2005}]{springel05}
    Springel V., et al., 2005, Nature, 435, 629

  \bibitem[]{} {Sobral, D., Best, P. N., Geach, J. E., Smail, Ian,
      Kurk, J., Cirasuolo, M., Casali, M., Ivison, R. J., Coppin, K.,
      Dalton, G. B., 2009, \mnras, 398, 75}


  \bibitem[\protect\citeauthoryear{Sunyaev \& Zel'dovich}{1970}]{sunyaev70}
    Sunyaev, R.A., \& Zel'dovich, Ya.B., 1970,  Astrophys. \& Space Science, 
    7, 3


\bibitem[]{}
{Wang. Y., 2008, JCAP 05, 021}

\bibitem[]{}
{Wang, Y., 2006, \apj, 647, 1}

  \bibitem[\protect\citeauthoryear{White}{2005}]{white05}
    White M., 2005, Astroparticle~Physics, 24, 334

\bibitem[]{}
{Yan}, L., {McCarthy}, P.~J., {Freudling}, W., {Teplitz}, H.~I., 
	{Malumuth}, E.~M., {Weymann}, R.~J., {Malkan}, M.~A., 1999, \apjl, 519, L47


  \bibitem[\protect\citeauthoryear{York et al.}{2000}]{york00}
    York D.G., et al., 2000, AJ, 120, 1579


\end{thebibliography}
